\documentclass[journal]{IEEEtran}
\usepackage{hyperref}
\usepackage{cite}
\usepackage{amsmath,amssymb,amsfonts}
\usepackage{algorithmic}
\usepackage{graphicx}
\usepackage{textcomp}
\usepackage{xcolor}

\ifCLASSINFOpdf
\else
\fi
\begin{document}
\title{Blind Localization of Early Room Reflections with Arbitrary Microphone Array}

\author{
    \IEEEauthorblockN{Yogev Hadadi\IEEEauthorrefmark{1}, Vladimir Tourbabin\IEEEauthorrefmark{2}, Zamir Ben-Hur\IEEEauthorrefmark{2}, David Lou Alon\IEEEauthorrefmark{2}, Boaz Rafaely\IEEEauthorrefmark{1}}
    \\\ 
    \\\
    \IEEEauthorblockA{\IEEEauthorrefmark{1} School of Electrical and Computer Engineering, Ben-Gurion University of the Negev, Beer-Sheva, Israel}
    \\\
    \IEEEauthorblockA{\IEEEauthorrefmark{2}Reality Labs Research @ Meta, Redmond, WA, USA}
}

\maketitle

\begin{abstract}
Blindly estimating the direction of arrival (DoA) of early room reflections without prior knowledge of the room impulse response or source signal is highly valuable in audio signal processing applications. The FF-PHALCOR (Frequency Focusing PHase ALigned CORrelation) method was recently developed for this purpose, extending the original PHALCOR method to work with arbitrary arrays rather than just spherical ones. Previous studies have provided only initial insights into its performance. This study offers a comprehensive analysis of the method's performance and limitations, examining how reflection characteristics such as delay, amplitude, and spatial density affect its effectiveness. The research also proposes improvements to overcome these limitations, enhancing detection quality and reducing false alarms. Additionally, the study examined how spatial perception is affected by generating room impulse responses using estimated reflection information. The findings suggest a perceptual advantage of the proposed approach over the baseline, with particularly high perceptual quality when using the spherical array with 32 microphones. However, the quality is somewhat reduced when using a semi-circular array with only 6 microphones
\end{abstract}

\begin{IEEEkeywords}
Direction-of-Arrival estimation, DoA estimation, Phase-Alignment, arbitrary array, early room reflections, sparse recovery
\end{IEEEkeywords}

% \fi
\IEEEpeerreviewmaketitle

\section{Introduction}

\IEEEPARstart{T}{he} estimation of early room reflections blindly, without prior knowledge of the room impulse response, has a role in various tasks, such as room geometry inference \cite{mabande2013room}, optimal beamforming \cite{javed2016spherical}, speech enhancement and dereverberation \cite{kowalczyk2017extraction}, \cite{peled2010method}, and source separation \cite{vincent2014blind}. Moreover, early reflections impact sound perception, enhancing speech intelligibility and creating a sense of listener envelopment, affecting the perceived source width, loudness, and distance \cite{catic2015role}, \cite{vorlander2007auralization}. 
As human-centered technologies evolve, enabling mobile users to interact seamlessly with natural wearable devices, there may be an increasing value for methods capable of estimating early room reflections from wearable arrays. Thus, the estimation of the Direction of Arrival (DoA) and delays of the early room reflections with arbitrary arrays can significantly advance emerging applications of audio signal processing \cite{pulkki2018parametric, coleman2017object, krim1996two}.

Spatial filtering\cite{665}, also known as beamforming, can be used as a blind method for estimating the directions of arrival of early reflections, which can be applied over arbitrary arrays\cite{ krishnaveni2013beamforming, 4445829, 9306271}. It effectively separates reflections from the direct sound, allowing for delay estimation through cross-correlation analysis \cite{mabande2013room}. However, this method may encounter challenges in practical scenarios with a high density of early reflections, as it has limitations in spatial resolution \cite{kuttruff2016room}. Methods based on subspace processing, such as MUSIC or ESPRIT\cite{krim1996two}, offer higher resolution possibilities \cite{sun2012localization, jo2019robust, ciuonzo2015performance, ciuonzo2017time} and can be employed with diverse array configurations \cite{6704870,157239}. However, these methods assume uncorrelated sources, which is unsuitable for early reflections since they are delayed replicas of the direct sound and highly correlated. Frequency smoothing \cite{5346492} can be applied to decorrelate source signals, but it is ineffective when reflections have similar delays. Additionally, these methods typically require the number of microphones to exceed the number of sources and reflections, which is often not the case in practical array setups. Another approach involves formulating the problem as an under-determined linear system and utilizing sparse recovery \cite{wu2012dereverberation}, but it is limited to detecting only a limited number of reflections. Alternative methods that model source signals as deterministic unknowns require complex non-linear optimization and may suffer from reduced spatial resolution \cite{sun2012localization,hu2018direction,van2004optimum,dmochowski2007generalized,dibiase2001robust,do2010srp}.

PHALCOR \cite{Shlomo2021Phalcor} tackles the constraints of previous techniques by leveraging the inherent coherence of the reflections. The phase-aligned transform is introduced to facilitate the separation of reflections in both temporal and spatial domains, thereby enabling the detection of more reflections than previous methods. However, the PHALCOR algorithm uses plane wave decomposition for obtaining Ambisonics signals, which limits its application to microphone arrays with specific geometry. Hence, this method is incompatible with general array configurations, including wearable. To extend the application of PHALCOR, a novel generalization approach called FF-PHALCOR was introduced by the authors of this paper \cite{hadadi2022towards,hadadi2023blind}, utilizing frequency focusing techniques \cite{BeitOn2020Focusing}. However, only initial findings were presented, demonstrating the capability of detecting multiple reflections successfully, but without a comprehensive analysis of performance and limitations. Furthermore, as indicated in \cite{hadadi2023blind}, incorporating information about the DoAs and delays of early room reflections can enhance spatial perception. However, this enhancement was demonstrated only with a spherical array.

This work conducts a more comprehensive investigation of the limitations of the FF-PHALCOR method using two different arrays in two ways. First, the performance and limitations are presented in terms of the detection quality and as a function of reflection characteristics such as delay and amplitude. Then, performance is assessed in terms of spatial enhancement in sound perception. This paper presents the following contributions: (1) Application of the method to a semi-circular array, including a comprehensive performance analysis (previous work studied only spherical arrays). (2) Understanding of the impact of reflection characteristics, such as delay, amplitude, and spatial density, on performance. (3) Algorithm improvement in the clustering stage to minimize both missed detections and false alarms in the estimation of reflections directions of arrival and delay.  (4) Demonstrating the application of the algorithm in improving spatial perception by incorporating reflections estimates into reproduced room impulse responses. 

This paper is structured as follows: Section II describes the mathematical notations and the system model. Section III describes the proposed algorithm. Section IV presents the results of the performance and limitations analysis using Monte Carlo simulation. Section V presents the results of the analysis in sound perception. Finally, the paper is concluded in Section VI.

\section{Mathematical Background}

\subsection{Notations:}

In this section, a brief overview of the notation employed in the paper is provided. Scalar variables are represented by lowercase letters (e.g $\tau$), vector variables are denoted by lowercase boldface letters (e.g $\mathbf{x}$), and matrix variables are indicated by uppercase boldface letters (e.g $\mathbf{V}$). $(\cdot)^{*}$, $(\cdot)^{T}$, $(\cdot)^{H}$, $\left|\left|\cdot\right|\right|$ denote the complex conjugate, transpose, conjugate transpose, and Euclidean norm, respectively. The set $S^2$ denotes the unit sphere. $\mathbb{E}\left[\cdot\right]$ denotes the expectation operation. The time and frequency domains are denoted by $t$ and $f$, respectively. 

\subsection{Signal Model}

The acoustic scenario involves a single source and an array of $Q$ microphones placed in a room. The anechoic signal in the frequency domain is denoted as $s(f)$, and each reflection is considered as attenuated and delayed replicas of the direct sound \cite{allen1979image}. Assuming the existence of the direct sound and $K$ early reflections, each is treated as a separate source $s_k(f)$, where $0\leq k\leq K$ as follows:

\begin{equation}
    s_k(f)=\alpha_k e^{-i2\pi f\tau_k}\emph{s}(f)
\end{equation}
The parameters $\alpha_k$ and $\tau_k$ denote the attenuation, and delay of the $k^{th}$ reflection, respectively. $s_0(f)$ denotes the direct sound with a normalized delay such that $\tau_0=0$, and normalized attenuation $\alpha_0=1$. The reflections are ordered in ascending order according to their delays, where $\tau_{k-1}\leq\tau_k$.

$\mathbf{s}(f) := [s_0(f),s_1(f),\dots,s_K(f)]^T$ is defined as the direct and $K$ early reflections source signal vector. In addition, $\mathbf{p}(f) := [p_1(f),p_2(f),\dots,p_Q(f)]^T$ is defined as the pressure signal captured at the microphone array where each element represents a single microphone, leading to the following array equation:

\begin{equation}\label{eq:pressure_on_mic}
   \mathbf{p}(f) = \mathbf{H}(f,\mathbf{\Omega})\mathbf{s}(f)+\mathbf{n}(f)
\end{equation}
$\mathbf{n}(f)$ represents the captured noise by the array and the reverberation part. $\mathbf{H}(f,\mathbf{\Omega})$ is the steering matrix with the frequency and the direction of arrivals of the reflections $\mathbf{\Omega}$ as arguments and is constructed as:
\begin{equation} \label{eq:H_definition}
    \mathbf{H}(f,\mathbf{\Omega}) := [\mathbf{h}(f,\Omega_0),\dots,\mathbf{h}(f,\Omega_K)]
\end{equation}
where $\Omega_k$ is the DoA of the k-th reflection. Each column $\mathbf{h}(f,\Omega_k)$ is a steering vector in a free field with respect to the $k^{th}$ reflection taking into account the acoustic effect of the array.

\section{The FF-PHALCOR Algorithm}

\begin{figure*}[t]
    \centering
    \includegraphics[width=\textwidth]{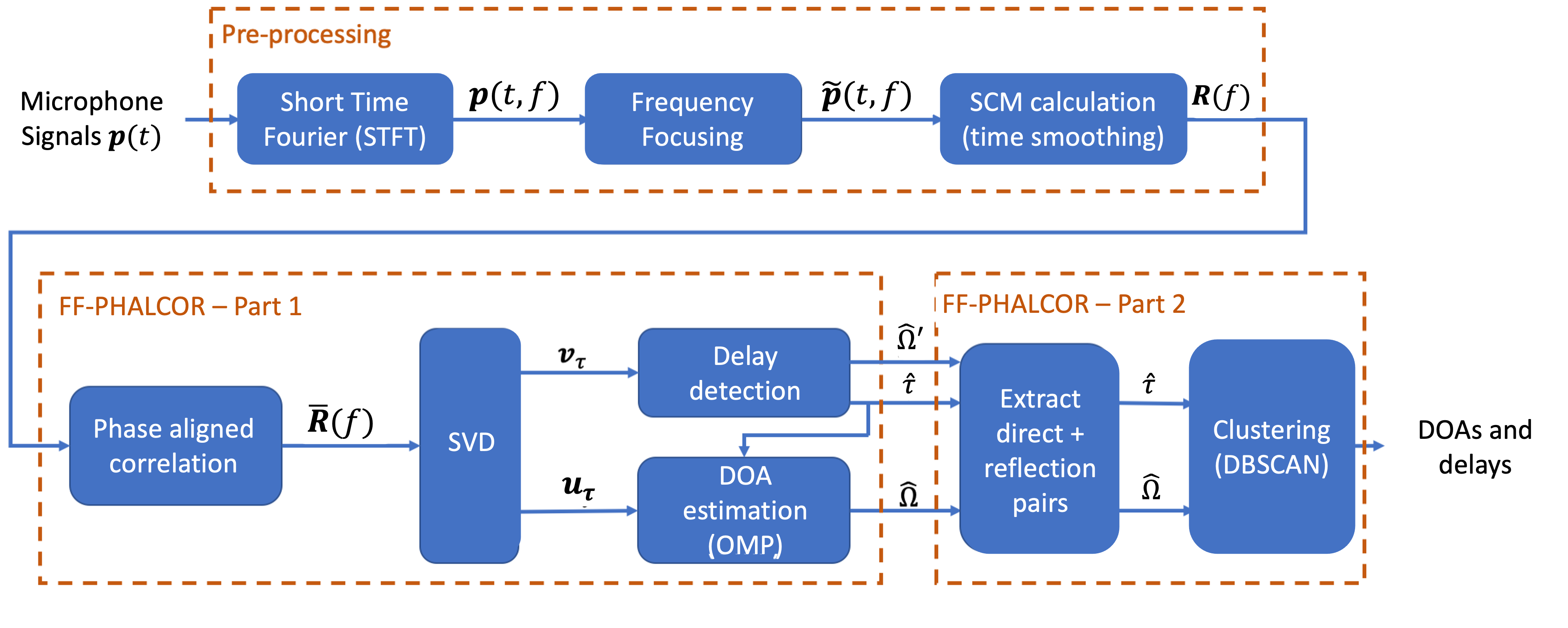}
    \caption{Block diagram of the algorithm}
    \label{fig:Algorithm22}
\end{figure*}

This section presents the FF-PHALCOR algorithm used in the work \cite{hadadi2022towards,hadadi2023blind}, based on the original PHALCOR algorithm \cite{Shlomo2021Phalcor}. The algorithm is shown in Fig. \ref{fig:Algorithm22}. In The PHALCOR algorithm, the spatial correlation matrix is first computed, and then the phase-alignment transform is performed. To apply the latter, a frequency-independent steering matrix must be obtained. In the original work \cite{Shlomo2021Phalcor}, a plane wave decomposition was initially used to address this limitation, which required a specific microphone array geometry. To overcome this restriction and allow application to arbitrary arrays, an extension of the algorithm was proposed \cite{hadadi2022towards} using frequency-focusing \cite{BeitOn2020Focusing} to achieve the frequency-independent steering matrix property.

\subsection{Frequency Focusing}

Eq. \eqref{eq:pressure_on_mic} establishes the array equation in the frequency domain. In most scenarios of arbitrary arrays, the steering matrix exhibits frequency dependence, posing challenges in applying the phase-alignment transform. To address this concern, a focusing process is introduced. First, the frequency domain is divided into bands with fixed bandwidth denoted as $B_w$. For each band, $f_0$ is defined as the center frequency of the band, and for each frequency in the band, a matrix $\mathbf{T}$ is defined to remove the frequency dependency by:

\begin{equation}\label{eq:define_T}
\mathbf{T}(f,f_0)\mathbf{H}(f,\mathbf{\Omega}) = \mathbf{H}(f_0,\mathbf{\Omega})    
\end{equation} 

The steering matrices $\mathbf{H}(f_0,\mathbf{\Omega})$ and $\mathbf{H}(f,\mathbf{\Omega})$ are constructed using Eq. \eqref{eq:H_definition}, where $f$ is a chosen frequency in the band. To obtain $\mathbf{T}(f,f_0)$, the pseudo-inverse of $\mathbf{H}(f,\mathbf{\Omega})$ is employed \cite{BeitOn2020Focusing}. This computation aims to minimize the mean-square error (MSE) between $\mathbf{H}(f,\Omega)$ and $\mathbf{H}(f_0,\Omega)$. Multiplying Eq. \eqref{eq:pressure_on_mic} with $\mathbf{T}(f,f_0)$, leads to:

\begin{equation}\label{eq:p_tilde_1}
    \tilde{\mathbf{p}}(f) = \mathbf{T}(f,f_0)\mathbf{p}(f) = \mathbf{H}(f_0,\mathbf{\Omega})\mathbf{s}(f) + \mathbf{T}(f,f_0)\mathbf{n}(f)
\end{equation}

\subsection{Phase Alignment Transform}

After the frequency-focusing step, the next stage involves applying the phase-alignment transform. This step entails computing the spatial correlation matrix (SCM) in the frequency domain by taking the expectation of the outer product between $\tilde{\mathbf{p}}$ and itself:

\begin{equation}\label{eq:define_R_p_tilde}
    \mathbf{R}(f) = \mathbb{E}[\tilde{\mathbf{p}}(f)\tilde{\mathbf{p}}(f)^H]
\end{equation}
Substituting Eq. \eqref{eq:p_tilde_1} into Eq. \eqref{eq:define_R_p_tilde}, and assuming uncorrelated signal and noise, leads to:

\begin{equation}\label{eq:define_R_FF}
    \mathbf{R}(f) = \mathbf{H}(f_0, \mathbf{\Omega})\mathbb{M}(f)\mathbf{H}(f_0,\mathbf{\Omega})^H + \mathbb{N}(f)
\end{equation}
where $\mathbb{N}(f)$ and $\mathbb{M}(f)$ are the expectation of the outer product of $\mathbf{T}(f,f_0)\mathbf{n}(f)$, and $\mathbf{s}(f)$, respectively. When the matrix $\mathbb{M}(f)$ is sparse, the Singular Value Decomposition (SVD) can be employed to extract the information about the signals. However, in practical scenarios, the matrix tends to be dense. To tackle this, the phase-alignment transform is introduced to enhance a single entry in the spatial correlation matrix, as follows \cite{Shlomo2021Phalcor}:

\begin{equation}\label{eq:define_R_bar}
    \overline{\mathbf{R}}(\tau,f) := \sum_{j=0}^{J_f-1} \omega_j  \mathbf{R}(f+j\Delta f)e^{i2\pi \tau j \Delta f}
\end{equation}
The parameter $J_f$ represents the total number of frequency points in the band, $\Delta f$ is the frequency resolution which satisfies $J_f\Delta f=B_w$, where $f$ is the first frequency in the band. The values $\omega_0,\dots,\omega_{J_f-1}$ are non-negative weights that are inversely proportional to $tr(\mathbf{R}(f_j))$\cite{Shlomo2021Phalcor}. As a result of the frequency-independent steering matrix and by substituting the formula from Eq. \eqref{eq:define_R_FF} into Eq. \eqref{eq:define_R_bar}:

\begin{equation}\label{eq:define_R_FF2}
    \overline{\mathbf{R}}(\tau,f) = \mathbf{H}(f_0, \mathbf{\Omega})\overline{\mathbb{M}}(\tau,f)\mathbf{H}(f_0,\mathbf{\Omega})^H + \overline{\mathbb{N}}(\tau,f)
\end{equation}
$\overline{\mathbb{M}}(\tau,f)$ and $\overline{\mathbb{N}}(\tau,f)$ are the results of applying the phase-alignment transform to matrix $\mathbb{M}(f)$ and $\mathbb{N}(f)$, respectively. In the original work \cite{Shlomo2021Phalcor}, it was shown that $\overline{\mathbb{M}}(\tau,f)$ is sparser then $\mathbb{M}(f)$. Also, it was shown that the maximum value of $\overline{\mathbb{M}}(\tau,f)$ occurs when $\tau = \tau_k-\tau_{k'}$, where $k'$ is chosen to represent the direct sound. Therefore, at this value of $\tau$, the transform enhances a single entry representing a single reflection, therefore providing a denoising effect. With a single entry in $\overline{\mathbb{M}}(\tau)$ enhanced, matrix $\overline{\mathbf{R}}(\tau)$ can be approximated by a rank-1 matrix which is denoted by $\overline{\mathbf{R}}_1(\tau)$ and obtained by truncating its SVD: 

\begin{equation}\label{eq:svd}
    \overline{\mathbf{R}}_1(\tau) = \sigma_{\tau}\mathbf{u}_{\tau}\mathbf{v}_{\tau}^H
\end{equation}
For each $\tau$, the left and right singular vectors are represented by $\mathbf{u}_{\tau}$ and $\mathbf{v}_{\tau}$, respectively, and the first singular value is denoted as $\sigma_{\tau}$. In cases where multiple reflections occur with the same delay, the right singular vector $\mathbf{v}_{\tau}$ closely approximates steering vector of the direct sound $\mathbf{h}(f_0,\Omega_0)$, and the left singular vector $\mathbf{u}_{\tau}$ becomes a linear combination of several $\mathbf{h}(f_0, \Omega_k)$ vectors, which correspond to the reflections at $\tau$.

\subsection{Detection}

After performing SVD, delay values that correspond to reflections are  determined by examining $\rho(\tau)$ and $\Omega'(\tau)$ \cite{Shlomo2021Phalcor}:

\begin{equation}\label{eq:rho}
    \rho(\tau) = \max_{\Omega'\in S^2}\frac{\left|\mathbf{h}(f_0, \Omega')^H \mathbf{v}_{\tau}\right|}{\|\mathbf{h}(f_0, \Omega')\|}
\end{equation} 
\begin{equation}\label{eq:Omega0_estimation}
    \hat{\Omega}'(\tau) = arg \max_{\Omega'\in S^2}\frac{\left|\mathbf{h}(f_0,\Omega')^H \mathbf{v}_{\tau}\right|}{\|\mathbf{h}(f_0, \Omega')\|}
\end{equation}
The Cauchy-Schwartz inequality ensures $\rho(\tau)\leq1$. Since $\mathbf{v}_{\tau}$ represents (up to a phase) the steering vector of the direct sound, $\hat{\Omega}'(\tau)$ is the estimated DoA for the direct sound. To obtain the DoA of reflections, the Orthogonal Matching Pursuit (OMP) algorithm \cite{cai2011orthogonal} is employed using vector $\mathbf{u}_{\tau}$, for each $\tau$  that lead to values of $\rho(\tau)$ exceeding an empirically determined threshold $\rho_{min}$, and $\hat{\Omega}'(\tau)$ being close to $\Omega_0$ up to an empirically determined threshold $\Omega_{th}$. 

\subsection{Clustering}\label{Sec::clustering}

The final step involves clustering the delay and DoA estimates to group them into dominant clusters representing reflections and to eliminate noise effects. This clustering is performed using the DBSCAN algorithm \cite{ester1996density}. 
In certain scenarios, the clustering algorithm generates clusters containing multiple adjacent reflections, resulting in the miss-detection of reflections. To address this issue, a sub-clustering method was proposed. The concept involves assessing whether a cluster can be divided into two or more clusters with centers located at a distance greater than a predefined threshold. The distance between the centers in the DoA-Delay map is calculated by computing the weighted Euclidean norm of the distances in each domain. The thresholds for splitting clusters were empirically determined.

\section{Simulation study}\label{sec:ExperimentalAnalysis}

Previous studies demonstrated the performance using specific acoustic conditions. In this paper, a comprehensive analysis was performed investigating the factors affecting misses and false alarms of the algorithm using a Monte-Carlo simulation that covers a wide range of conditions. The setup and methodology are presented in the next sections, followed by the results and improvements.

\begin{table}
\caption{Dimensions, $T_{60}$, and the average number of reflections within the first 20ms  of the simulated rooms}
\begin{center}
\begin{tabular}{|c|c|c|c|c|}
\hline
 & Dim $\left[m\right]^3$ &  $T_{60}$ [s] & \#Reflections \\
\hline
1 & $12\times9\times5$ & 1.22 & 7.9 \\
\hline
2 & $10\times7\times4$ & 0.99 & 12 \\
\hline
3 & $9\times5\times3$ & 0.89 & 18.9 \\
\hline
4 & $6\times4\times3$ & 0.62 & 25.6  \\
\hline

\end{tabular}
\label{tb:Rooms_table}
\end{center}
\end{table}

\subsection{Setup} \label{sec:simSetup}
The Monte Carlo simulation contained $300$ simulated scenes, each containing a simulated room chosen randomly but with equal probability from Table \ref{tb:Rooms_table}. A point source and an array were positioned in each room at a minimum distance of $1.2\,$m from the walls, with varying distances between them ranging from $0.7\,$m to $1.7\,$m. Additionally, the positions of the source and array were perturbed by a random distance in the range $\left[-\frac{1}{2},\frac{1}{2}\right]\,$m. Simulation conditions were controlled to achieve a DRR (Direct-to-Reverberant Ratio) value within the range of $\left[-10, 10\right]\,$dB. Two arrays were simulated - a spherical array similar to the em$32$\cite{acoustics2013em32} and a semi-circular array, which contains $6\,$microphones uniformly distributed over half a circle in the equator of an open sphere. The room impulse response is calculated using the image method \cite{allen1979image}. Anechoic speech signals, used as source signals,  were taken from the TSP Database\cite{kabal2002tsp} with a length of $2.66$ to $3.12\,$seconds and included two male and two female speakers, sampled at $16\,$kHz. The simulated RIR were then used to compute the sound field around the arrays in the spherical harmonics domain, including terms up to order $N=8$ as in \cite{rafaely2015fundamentals}. 

\begin{figure}[t]
    \begin{center}
         \centering
        
             \includegraphics[width=0.95\columnwidth]{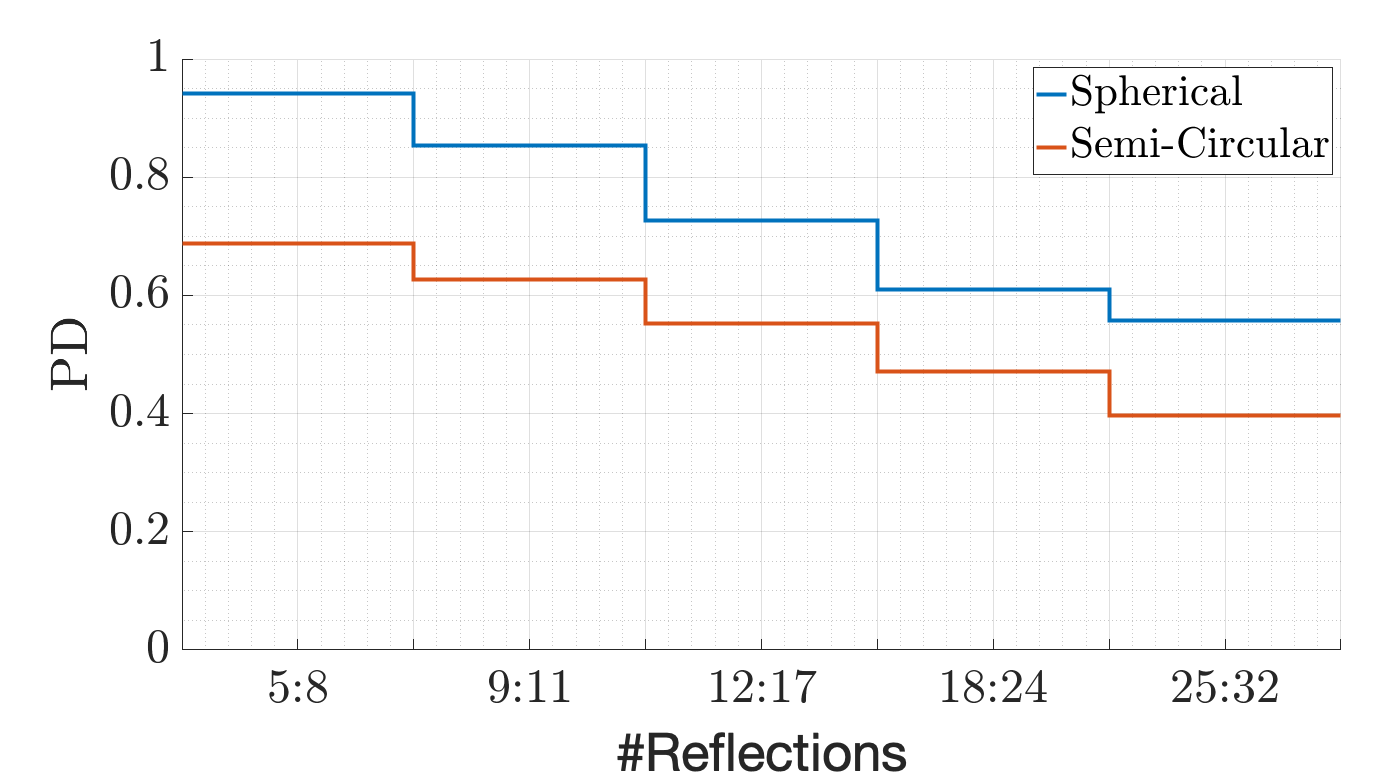} 
             
             (a)

             \includegraphics[width=0.95\columnwidth]{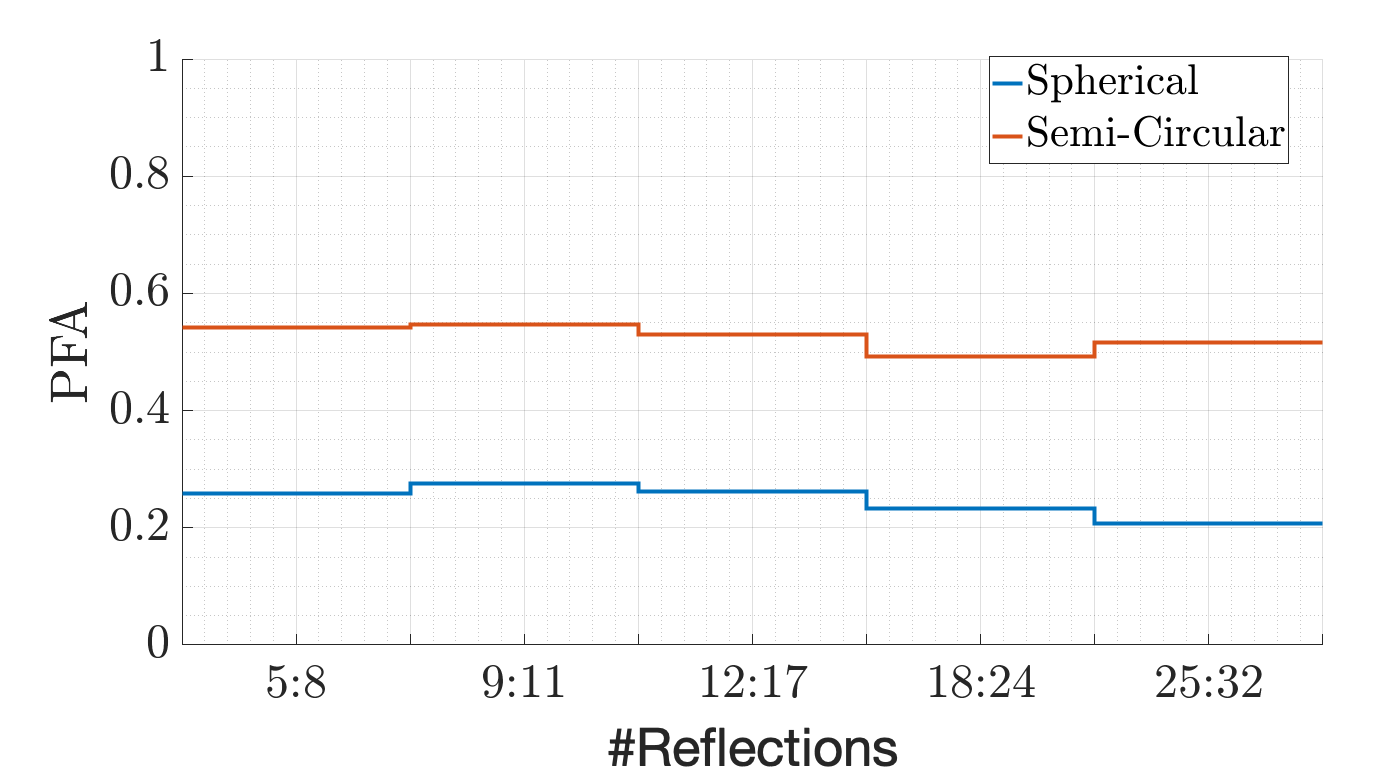}
             
             (b)
        
        \caption{(a) Mean probability of Detection $P_D$ and (b) Mean probability of False Alarms $P_{FA}$, across simulations, as a function of the number of reflections in the ground truth within the first $20\,$ms. The X-axis indicates the count of reflections in $5$ groups, and the Y-axis shows the values of (a) $P_D$ and (b) $P_{FA}$. The outcomes for the semi-circular and em32 arrays are illustrated by the orange and blue lines, respectively.}
        \label{fig:PD_PFA_semi_circular_em32}
    \end{center}

\end{figure}

\subsection{Methodology}

A Short-Time Fourier Transform (STFT) is applied to the microphone signals using a Hanning window of $150\,$ms, a sampling frequency of $16\,$kHz, and a $75\%\,$ overlap. The operating frequency range used for processing spans from $500\,$Hz to $5000\,$Hz and is divided into bands of $2\,$kHz each. This division is achieved by creating $11$ bands starting at 500 Hz, each with an overlap of approximately 1740 Hz. %% not passing 5k so last band starts at 3k.
The focusing matrix, as detailed in Eq. \eqref{eq:define_T}, was derived from a simulated steering matrix, which was constructed using $900$ DoAs sampled via the Fliege-Maier method \cite{fliege1996two}. The hyper-parameters, as specified in the original paper \cite{Shlomo2021Phalcor}, are set as follows: $\rho_{min}=0.9$, $\epsilon_u = 0.63$, $\Omega_{th} = 10^{\circ}$, $S_{max} = 3$, $\gamma_{\Omega} = 15^{\circ}$, $\gamma_{\tau} = 0.3\,$ms, and the density threshold to $0.05$. Additionally, the threshold for the sub-clustering algorithm was selected to be $0.25$.

For performance evaluation, a reflection is identified as a true positive when it matches the ground truth (determined from the image method) within the allowed tolerances of $0.5\,$ms in delay, and $15^{\circ}$ in DoA. The Probability of Detection ($P_{D}$) and the Probability of False Alarms ($P_{FA}$) for each tested scenario are defined as follow: 

\begin{equation}\label{eq:PD}
    P_D := \frac{\#\;true\;positive\;detections}{\#reflections\;in\;the\;ground\;truth}
\end{equation}

\begin{equation}\label{eq:PFA}
    P_{FA} := \frac{\#\;false\;positive\;detections}{\#detected\;reflections}
\end{equation}
Furthermore, the analysis extends to calculate the mean and standard deviation across all Monte Carlo simulations for the examined parameters (e.g., the number of reflections within the first $20\,$ms, delay, and amplitude variations). Additionally, the Probability of Miss-detection ($P_{M}$) is determined, adhering to the relation $P_M = 1 - P_D$.

\subsection{Overall performance}

Fig. \ref{fig:PD_PFA_semi_circular_em32} presents the results of the Monte Carlo simulation for the two metrics, $P_D$ and $P_{FA}$, as a function of the number of reflections. It demonstrates a clear trend where an increase in the number of reflections within the initial $20\,$ms leads to a reduction in $P_D$. Furthermore, the analysis reveals that the $P_D$ associated with the semi-circular array significantly falls below that of the em32 array, whereas the $P_{FA}$ for the semi-circular array significantly exceeds that of the em32. A comprehensive analysis was undertaken to provide some insight into these results. Factors such as reflections amplitude, reflections delay, clustering of reflections, and ambiguity in reflections DoA estimations were thoroughly investigated

\begin{figure}[h!]
    \begin{center}
         \centering
         \includegraphics[width=0.95\columnwidth]{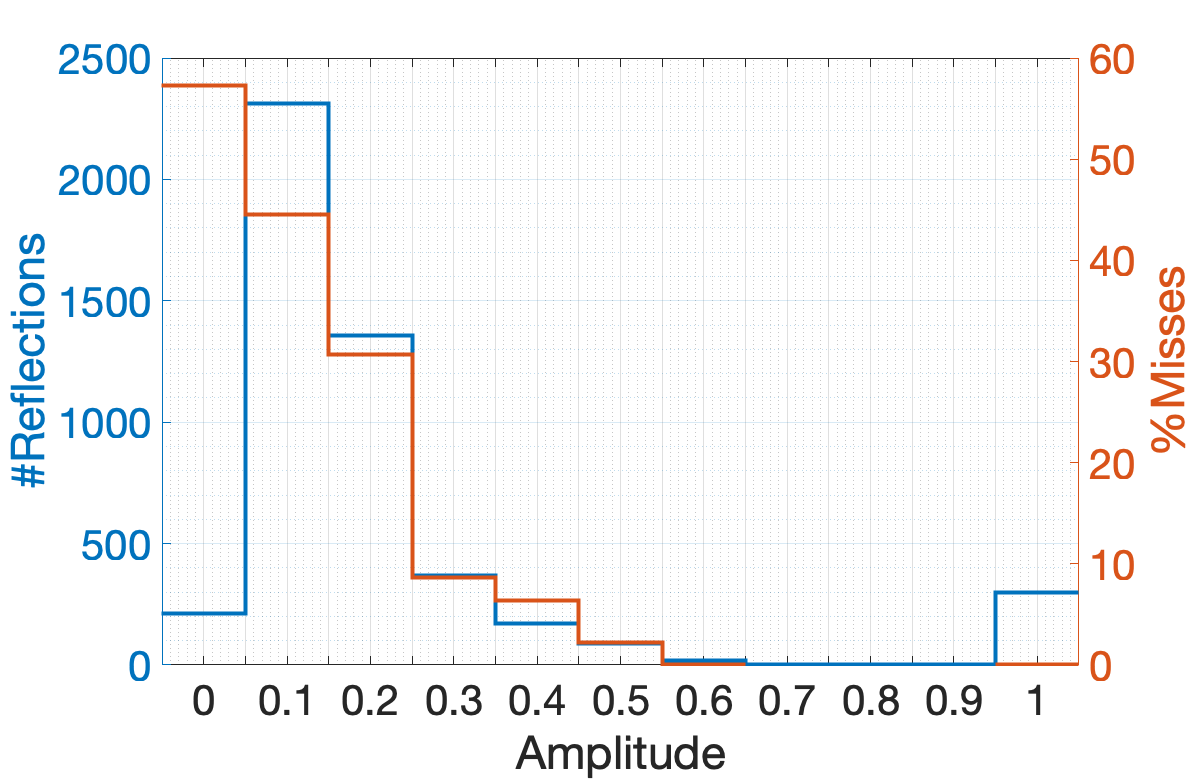}
         (a)
          \includegraphics[width=0.95\columnwidth]{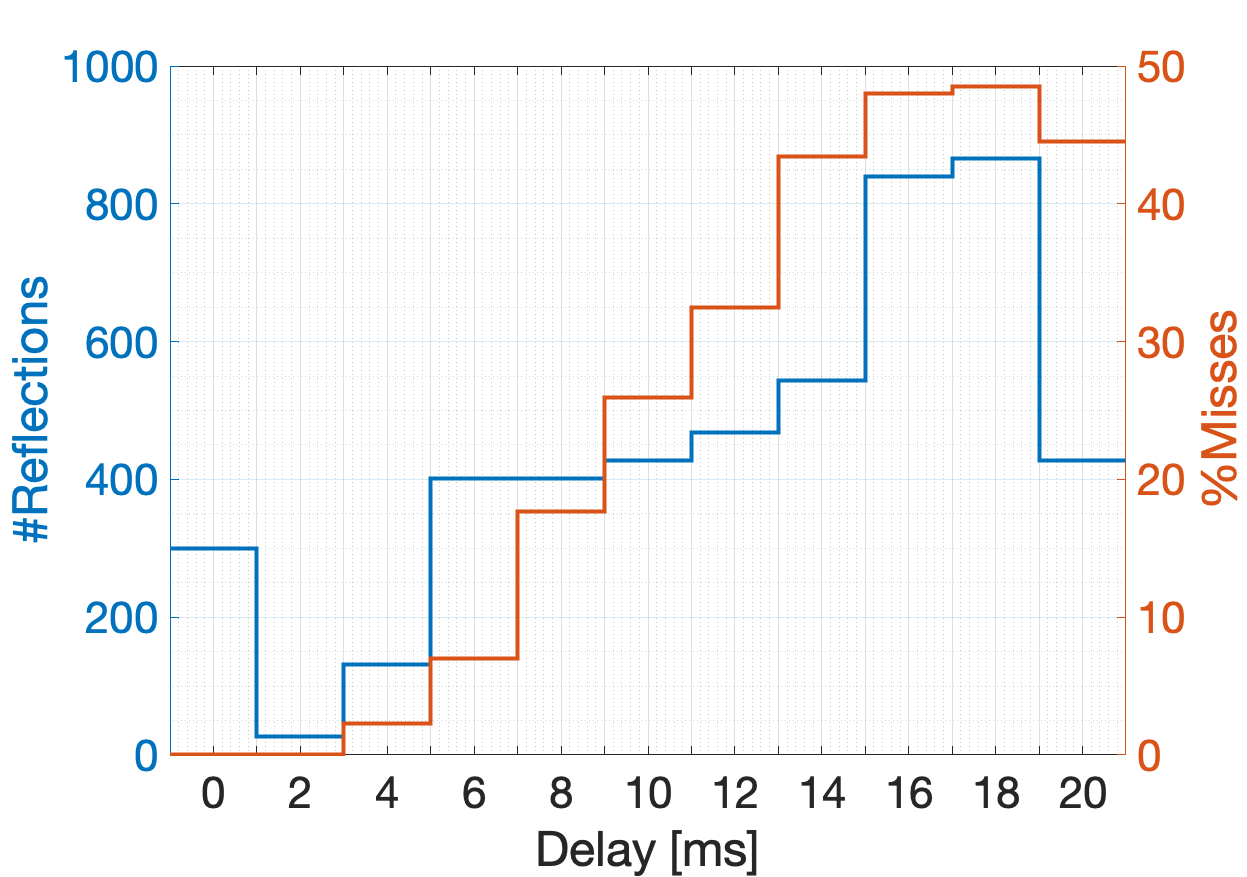}
          (b)
        \caption{(a) The probability of misses ($P_M$) and the count of reflections across varying ranges of reflections amplitudes, and (b) Similar to (a), with reflection delay replacing amplitude.}
        \label{fig:Misses_vs_Amp_Delay}
    \end{center}

    \begin{center}
         \centering
         \includegraphics[width=0.95\columnwidth]{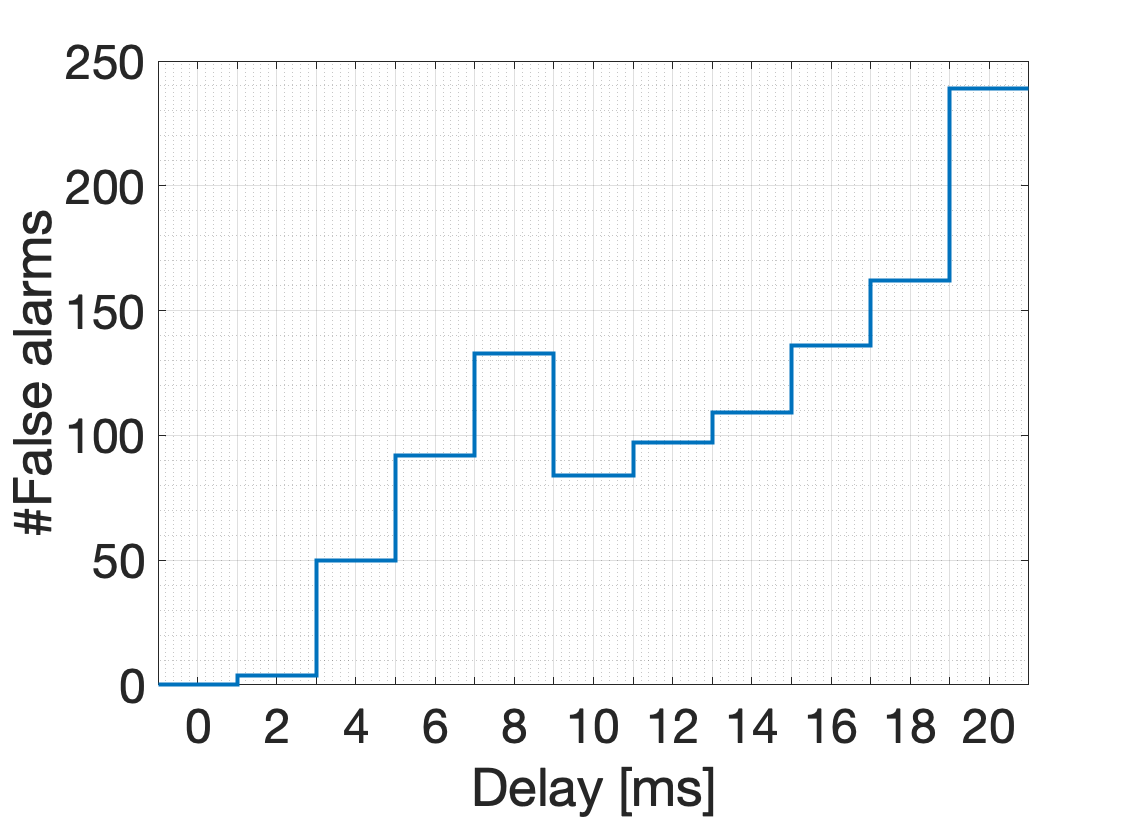}
  
        \caption{The graph shows false alarms plotted against the corresponding delay of the estimates. The blue line represents the count of reflections at various delay intervals.}
        \label{fig:FAs_vs_Delay}
    \end{center}

\end{figure}

\subsection{The effect of reflections delay and amplitude}

The aim of this section is to gain insight into the impact of the amplitude and delay of reflections, on the Probability of Detection ($P_D$), Probability of Misses ($P_M$), and the Probability of False Alarm ($P_{FA}$). Note that $P_D = 1 - P_M$, rendering the presentation of both probabilities redundant. In Fig. \ref{fig:Misses_vs_Amp_Delay}(a) reflection amplitude values were normalized so that the direct sound amplitude equals $1$, establishing a baseline for comparison. The blue line indicates the count of reflections at each amplitude level range, and the orange line shows the rate of misses per amplitude, normalized against total occurrences. Fig. \ref{fig:Misses_vs_Amp_Delay}(a) clearly shows that both the number of reflections and $P_M$ decrease in a similar manner with increasing reflection amplitude. This clearly shows that reflections with high amplitudes, i.e. $0.4$ with respect to the direct sound or higher, are easier to detect. It should be noted that in the amplitude range from approximately 0.7 to 0.9, there were no reflections, and so in the blue curve remained at $0$.

In Fig. \ref{fig:Misses_vs_Amp_Delay}(b) reflection delay values were normalized so that the direct sound delay equals $0$. The blue line indicates the count of reflections at each delay interval in milliseconds, with the orange line mapping out the miss rate relative to the total reflections for those intervals. Fig. \ref{fig:Misses_vs_Amp_Delay}(b) clearly shows that both the number of reflections and $P_M$ decrease in a similar manner with decreasing reflection delay. This clearly shows that reflections that appear later, i.e. $10\,$ms with respect to the direct sound, are harder to detect. In both Fig. \ref{fig:Misses_vs_Amp_Delay}(a) and \ref{fig:Misses_vs_Amp_Delay}(a), it seems like there is a correlation between $P_M$ and amplitude and delay. It might suggest that the reason why the misses are high is not due to the low amplitude or high delay; it is just due to the number of reflections, which makes the detection harder.

Another interesting insight can be gained by studying the effect of delay on the incidence of false alarms. In Fig.\ref{fig:FAs_vs_Delay}, the blue line indicates the count of false alarms at each delay interval in milliseconds. The delays are normalized as before. Fig. \ref{fig:FAs_vs_Delay} clearly shows that an increase in delay is associated with a rise in false alarms. This observation underscores the significance of delay as a factor in the increased probability of false alarms. 

Observations indicate that at lower delay values, the probability of misses ($P_M$) and the number of false alarms remain low. This suggests that reflections can be effectively detected up to a certain delay threshold, beyond which detection efficiency declines.

\subsection{The effect of local reflections clustering}\label{sec:ClusteringEffect}

\begin{figure}[t]
    \begin{center}
         \centering
         \includegraphics[width=1\columnwidth]{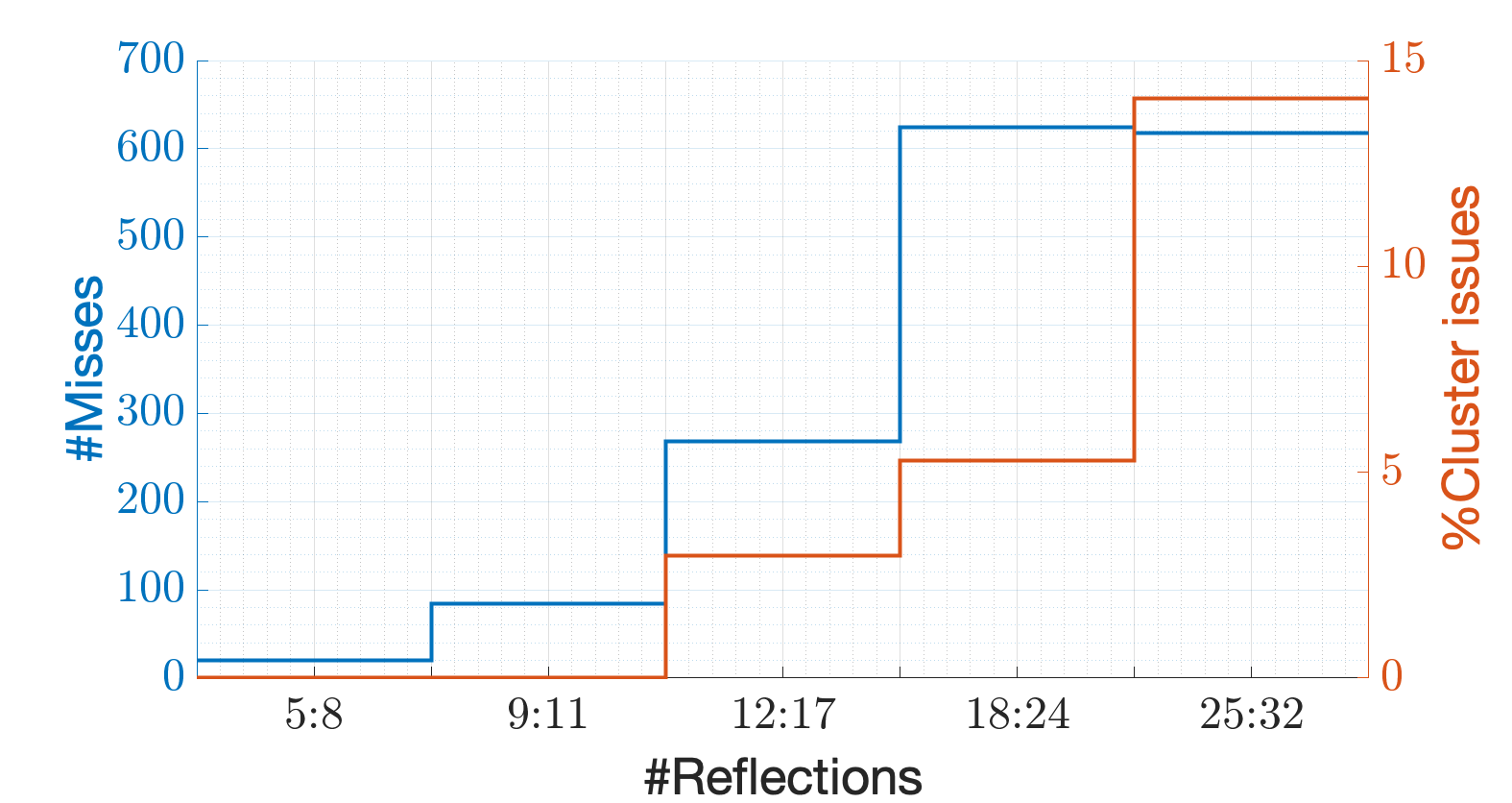}
  
        \caption{Misses resulting from the clustering of multiple reflections into a single cluster. The blue line indicates the total misses, categorized by the range of the number of reflections shown on the X-axis. The orange line illustrates the proportion of these misses as a percentage of the overall misses.}
        \label{fig:Misses_vs_Clustering}
    \end{center}

\end{figure}

The aim of this section is to investigate the effect of clustering algorithms on the incidence of detection misses. The aim is to understand whether the algorithm tends to combine multiple reflections into single clusters, thereby categorizing them as a single reflection, leading to misses. In Fig. \ref{fig:Misses_vs_Clustering}, the blue line indicates the total number of missed detections, categorized by the number of reflections within the first $20\,$ms of the ground truth. The red line shows the percentage of reflections that were classified in a single cluster together with other reflections. Fig. \ref{fig:Misses_vs_Clustering} clearly shows a high number of reflections in the chosen time-frame, i.e. $28$-$32$, leads to a higher rate of misses due to multi-reflection clustering. In the worst case, up to $15\%$ of misses were due to this problem. This behavior seems reasonable - with more reflection, their density in time and space increases, leading to more reflections being clustered together.

\begin{figure}[t]
    \begin{center}
         \centering
         \includegraphics[width=0.95\columnwidth]{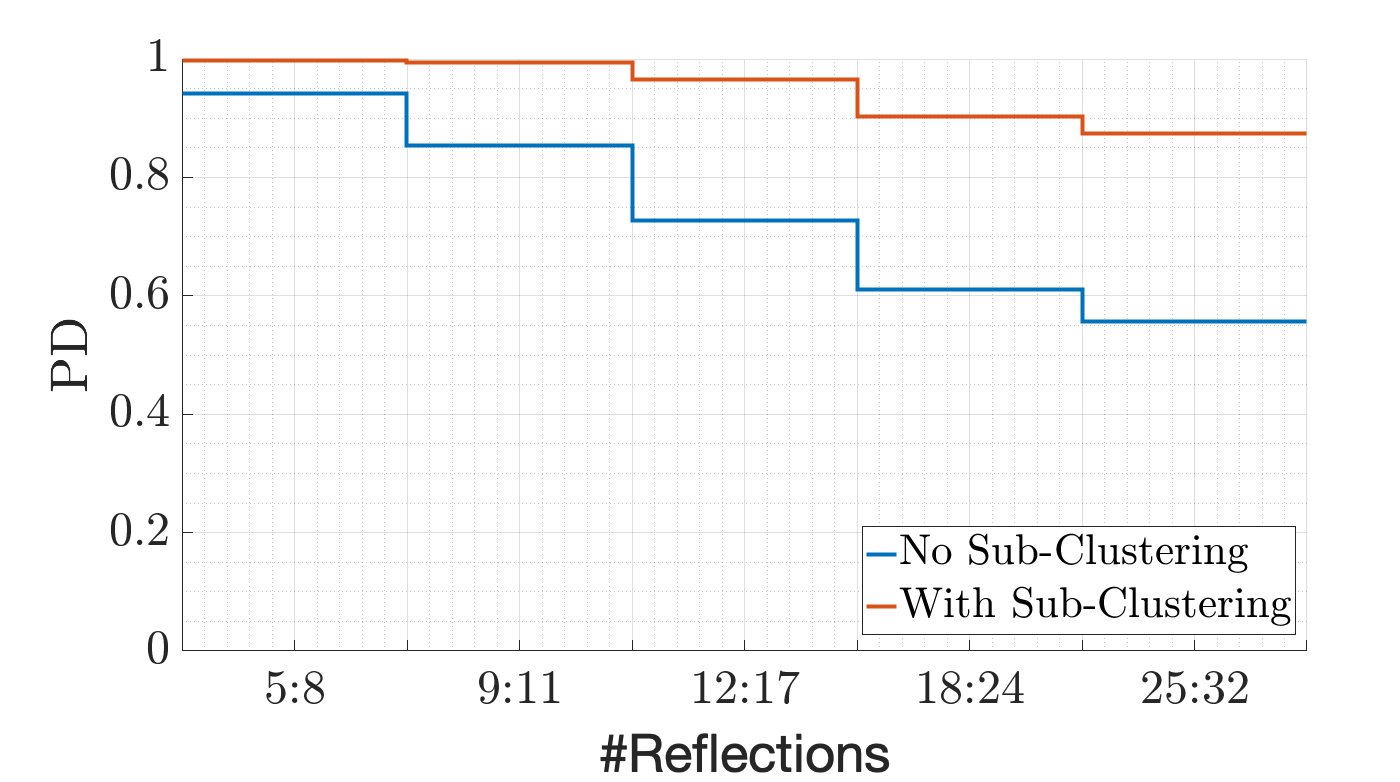}
  
        \caption{Mean Probability of Detection ($P_D$) as a function of the range of reflections in the ground truth. The blue line shows $P_D$ without the sub-clustering algorithm, while the orange line shows $P_D$ with the sub-clustering algorithm applied.}
        \label{fig:Kmeans}
    \end{center}

\end{figure}

To mitigate this problem, it is proposed to implement a sub-clustering mechanism to separate these larger clusters into sub-clusters. This sub-clustering was implemented using the k-means algorithm \cite{steinley2006k} as noted in Section \ref{Sec::clustering}, and applied separately to the data of individual clusters. Fig. \ref{fig:Kmeans} investigates the effectiveness of this approach, showing the probability of detection against the total number of reflections within the first $20\,$ms of the ground truth. The blue line shows the results for the original algorithm, while the red line shows the results of the modified algorithm (using sub-clustering). It can be seen that employing a sub-clustering algorithm notably enhances detection probabilities in contrast to the baseline algorithm. It was also verified that this improvement in $P_D$ did not come at the expense of any degradation in misses $P_M$.

\subsection{The effect of ambiguity in array steering}\label{sec:AmbiguityEffect}
\begin{figure}[t]
    \begin{center}
         \centering
         \includegraphics[width=0.95\columnwidth]{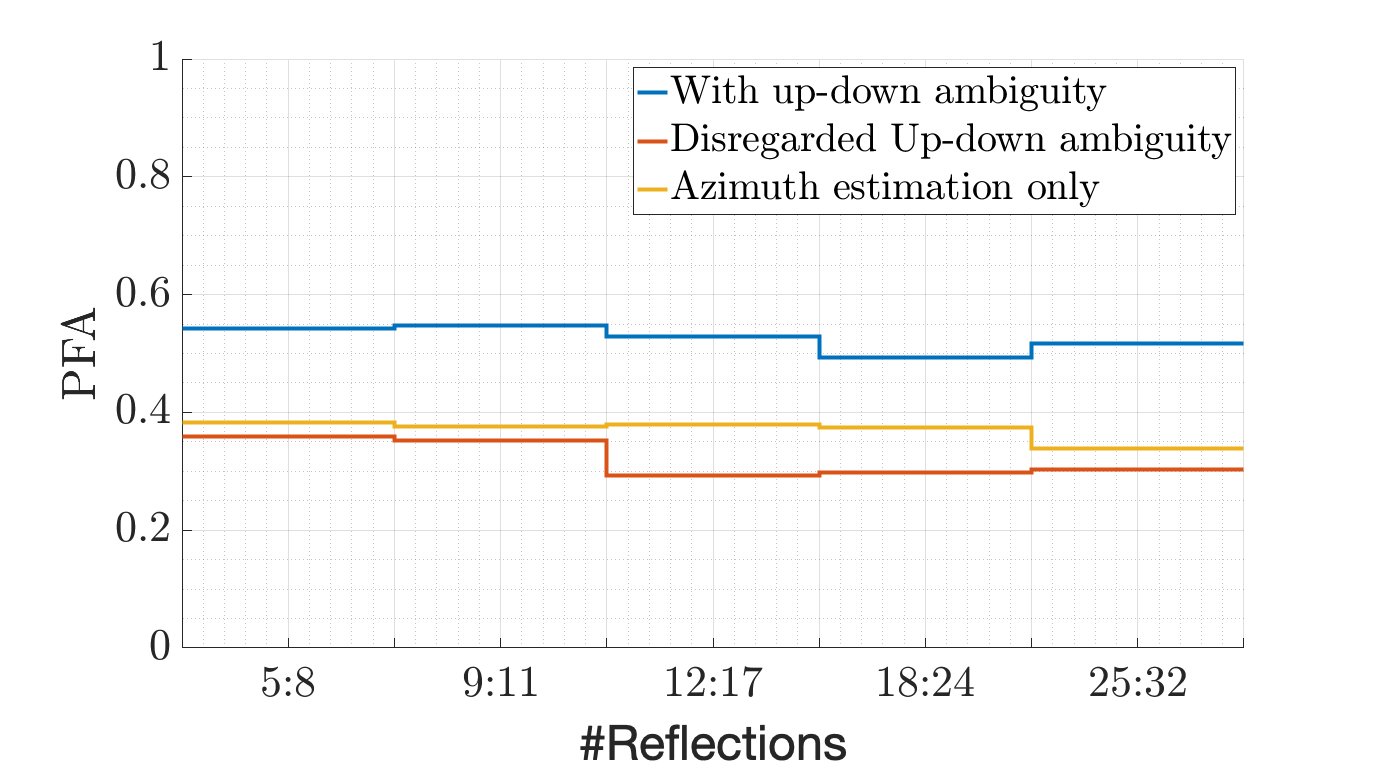}
  
        \caption{Mean Probability of False Alarms ($P_{FA}$) for the semi-circular array. The blue line shows probabilities when considering both elevation and azimuth, the red line shows the effect of resolving elevation ambiguity, and the orange line indicates the scenario when only azimuth is considered, disregarding elevation ambiguity. These probabilities are plotted against the number of reflections.}
        \label{fig:FA_ambiguity}
    \end{center}

\end{figure}

The aim of this section is to explore the influence of directional ambiguity in array steering function on the incidence of false alarms. In Fig. \ref{fig:PD_PFA_semi_circular_em32}(b), the semi-circular array is shown to have a higher probability of false alarms compared to the em$32$ array. This difference is linked to the design of the semi-circular array, which features $6$ microphones arranged along a horizontal semi-circle on an open sphere. Such an arrangement creates an ambiguity in determining whether the origin of a reflection is the upper or the lower half-space. Fig. \ref{fig:FA_ambiguity} shows the probability of false alarms against the number of reflections within the first $20\,$ms of the ground truth. The blue line shows the original $P_{FA}$, while the red line shows the $P_{FA}$ after eliminating the effect of this ambiguity by correcting lower vs. upper half-plane errors. The yellow line is the same as the red line, but when considering only the azimuth estimation errors. Fig. \ref{fig:FA_ambiguity} shows a significant reduction in $P_{FA}$ when applying ambiguity corrections in comparison to the original estimations, which may indicate that the up-down ambiguity significantly influences the false alarms rate. Following this observation, and estimating only the azimuth direction, disregarding the elevation ambiguity, leads to improved $P_{FA}$, thereby enhancing detection accuracy. 

\subsection{Summary}
The results and analysis of this section reveal that the FF-PHALCOR algorithm's performance is closely tied to the characteristics of the reflections it detects. The Monte Carlo simulation demonstrated that an increase in the number of reflections within the first $20\,$ms leads to a reduction in the $P_D$, particularly for the semi-circular array, which showed lower $P_D$ and higher $P_{FA}$ compared to the spherical array. This was attributed to the semi-circular array's fewer microphones and limited spatial coverage, as well as the directional ambiguity inherent in its design, which caused difficulties in accurately distinguishing between reflections from the upper and lower hemispheres. The analysis further highlighted that reflections with higher amplitudes were easier to detect, while those with lower delays were less likely to result in false alarms. Additionally, the study examined the effects of clustering reflections, which can lead to missed detections, and proposed improvements in the algorithm to address this issue. The analysis also found that by focusing only on azimuth angles, the false alarm rate in the semi-circular array could be significantly reduced. These insights suggest that addressing the ambiguity effect in the semi-circular array is crucial to enhance its practical applicability. Overall, the chapter provides valuable insights into the algorithm's limitations and suggests areas for refinement to improve its performance.

\section{Listening Test}

In the previous section, the performance of the algorithm was tested, demonstrating that it effectively detects early reflections. The performance was assessed using the  $P_d$, $P_{FA}$, and $P_M$ metrics, and the results were compared to the ground truth. While this analysis is useful, previous studies \cite{kitic2023blind,rafaely2022spatial,hadadi2023blind} have established the importance of early reflections in spatial perception. To assess the application of FF-PHALCOR within this context, a listening test based on the MUSHRA method (Multiple Stimuli with Hidden Reference and Anchor) \cite{series2014method} was conducted. Performance was evaluated using both the spherical and the semi-circular arrays, with the latter representing a more practical array alternative with a fewer microphones, therefore providing fewer true detections and more false alarms. 

\subsection{Setup}\label{sec:Setup}

A simulated setup was generated, consisting of a shoe-box room with dimensions of $12\times7\times5$ meters, and $T_{60} = 0.57\,$s. A speaker simulated by a point source is positioned at $[x,y,z] = [5.5, 1.2, 1.7]$, and the array is positioned at $[x,y,z] = [3.5, 3, 1.7]$. The elevation of the direct sound was set at the horizontal plane, and the azimuth at $42^\circ$ to the right. Two different arrays were simulated: a $32$-channel spherical microphone array and a $6$-channel semi-circular array, as described in the previous chapter. A $2.5\,$second segment from the TSP Speech Database \cite{kabal2002tsp} was used as the source signal. The head-related transfer function (HRTF) applied in the experiment is derived from the Neumann KU-$100$ \cite{bernschutz2013spherical}.

\subsection{Methodology}

The test consists of four signals: the reference signal, a synthesized anchor signal, and two estimated signals. One estimated signal is obtained from the spherical array using the sub-clustering technique described in Section \ref{sec:ClusteringEffect}, while the other is obtained from the semi-circular array using the azimuth-only method to decrease false alarms, as explained in Section \ref{sec:AmbiguityEffect}. 

\subsubsection{The Reference Signal}
The reference signal is generated by applying the image method to compute the room impulse response, simulating each reflection as an attenuated and delayed copy of the direct sound, based on the room dimensions and the walls' reflection coefficients \cite{allen1979image}. To fully capture the impact of early reflections and eliminate the effect of the reverberant part, the room impulse response was truncated to include only the direct sound and early reflections with arrival time not longer than $20\,$ms after the direct sound. While it is acknowledged that the truncated response does not represent a practical room response, it is used in this study to highlight the effect of the early reflections. Further research is proposed for extending this study to full room responses. The truncated room impulse response was then convolved with the head-related impulse response (HRIR) to create the Binaural Room Impulse Response (BRIR) \cite{rafaely2010interaural}, which was subsequently convolved with the anechoic source signal.

\subsubsection{The Synthesized Signal} The synthesized signal was used as a baseline or anchor for comparison, allowing us to evaluate the effectiveness of the algorithm in enhancing spatial perception.
The synthesized anchor signal is generated assuming the actual RIR is not available, but some information about the room is known with limited accuracy. It assumed that - the DoA of the direct sound is known accurately; $T_{60}$ is known approximately, computed from non-ideal estimates of the room volume and absorption; early reflection timing and directions are unknown and generated statistically; and reflection amplitudes are estimated from RIR decay imposed by $T_{60}$. This model relies on the formula for calculating the number of reflections in a shoe-box room over time $t$
\cite{kuttruff2016room}, as follows:
\begin{equation}
    N(t) = \frac{4\pi(ct)^3}{3V}
\end{equation}
where $N(t)$ represents the average number of reflections at time $t$, $c$ is the speed of sound, and $V$ is the non-ideal estimated volume of the room. The room volume was estimated with an error margin of up to $30\,m^3$, which can be attributed to a $0.5\,m$ error in estimating the room dimensions, resulting in an estimated volume of $390\,m^3$. The time axis was divided into $1\,$ms intervals, and for each interval $[t_i, t_{i+1}]$, the average number of reflections was calculated as follows:

\begin{equation}
    N_i = N(t_{i+1}) - N(t_i)
\end{equation}
Then, for each interval $[t_i,t_{i+1}]$, $N_i$ reflections are generated with delays evenly distributed within the interval and random directions of arrival uniformly sampled over $[0,\pi]$ for elevation and $[-\pi,\pi]$ for azimuth. To estimate the amplitudes of the early reflections, the $T_{60}$ was computed using the Schroeder integral \cite{schroeder1965new}, and the amplitudes were set to fit the energy decay curve. Then the synthesized room impulse response was convolved with the known head-related impulse response (HRIR) and the anechoic source signal, following the same process used for the reference signal.

\subsubsection{The Estimated Signals} The estimated signals are generated in a similar manner to the synthesized signals. The results of the FF-PHALCOR algorithm, which provide a set of delays and DoAs, as described in Section \ref{Sec::clustering} are used to replace the statistical estimates of DoA and delay used in the synthesized signal. The amplitudes of the early reflections are computed using the energy decay curve in the same manner as the synthesized signal.

The performance of the algorithm was evaluated using the metrics $P_D$ and $P_{FA}$ as previously defined in Eq. \eqref{eq:PD} and Eq. \eqref{eq:PFA}, respectively. For the spherical array, the sub-clustering method was applied to enhance the probability of detection. Following the analysis, the room impulse responses were generated and processed by convolution with the HRIR and the anechoic source signal. The spherical array achieved 
$P_D=1$ and the semi-circular array achieved 
$P_D=0.85$, with both arrays achieving 
$P_{FA}=0.4$
%Two different arrays were considered to obtain the estimated signals: the spherical array and the semi-circular array, as described in Section \ref{sec:simSetup}. The performance of the algorithm was evaluated using the metrics $P_D$ and $P_{FA}$ as defined in Eq. \eqref{eq:PD} and Eq. \eqref{eq:PFA}, respectively. These metrics were utilized in the experimental analysis described in Section \ref{sec:ExperimentalAnalysis}. For the spherical array, the sub-clustering method was utilized, as detailed in Sections \ref{Sec::clustering} and \ref{sec:ClusteringEffect}, to enhance the probability of detection. Finally, the room impulse responses were generated, and then convolved with the HRIR and the anechoic source signal. 
The synthetic and estimated signals were subsequently equalized in power and delay to match the reference signal. Finally, headphone equalization was applied to ensure consistent playback quality.

Sixteen participants with self-reported normal hearing were recruited for the listening test. The experiment utilized a single MUSHRA screen, where all four signals were presented through MATLAB's audio player (MATLAB R2022b). Participants rated the similarity to the reference with respect to overall quality on a scale from $0$ to $100$. To ensure comprehension of the instructions and familiarity with the signals, participants completed a training task before providing their ratings.

\subsection{Results}
\begin{figure}[t]
    \begin{center}
         \centering
         \includegraphics[width=0.95\columnwidth]{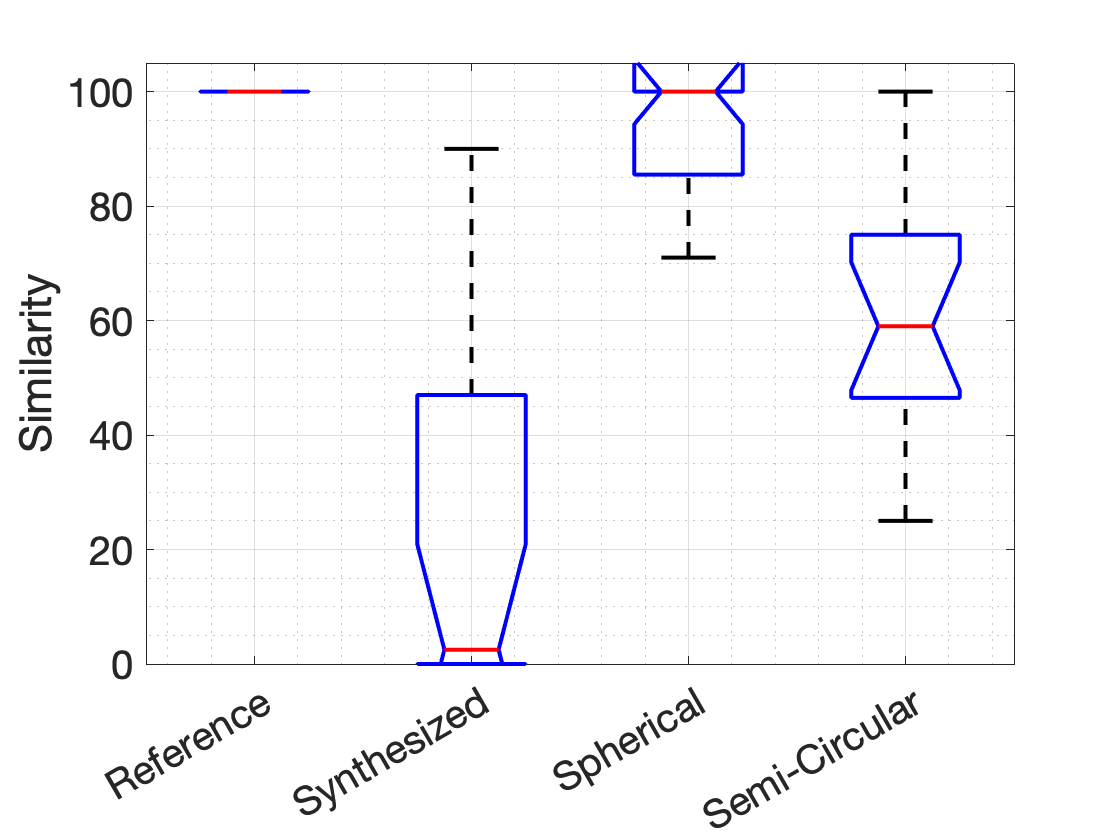}
  
        \caption{Similarity to the reference with respect to overall quality, displayed in a box plot, showing four signals: Reference, Synthesized, Estimated using Spherical array, and Estimated using Semi-Circular array. In the plot, the red line indicates the median, the box edges denote the $25$th and $75$th percentiles, and the whiskers represent the range of values, excluding outliers. The notches show the $95\%$ confidence interval.}
        \label{fig:BoxPlotSignals}
    \end{center}

\end{figure}

%As mentioned in the previous section, the algorithm's performance was first evaluated using the $P_D$ and $P_{FA}$ metrics. For the spherical array, the algorithm achieved $P_D = 1$ and $P_{FA} = 0.4$. For the semi-circular array, only the azimuth angles were considered, as this method proved more accurate in reducing false alarms, achieving $P_D = 0.85$ and $P_{FA} = 0.4$. 

Fig. \ref{fig:BoxPlotSignals} displays the results of the test conducted in this study. The four boxes represent the evaluated signals, with the y-axis indicating the similarity to the reference signal, where $100$ denotes perfect similarity and $0$ indicates lowest similarity. The red line within each box represents the median, while the blue lines denote the $25$th and $75$th percentiles. The whiskers show the range of values, excluding outliers. As illustrated, the median for the reference signal is at $100$ with no variance, indicating perfect detection every time. The spherical array results closely follow the reference, followed by the semi-circular array, with the synthesized signal receiving the lowest scores overall. 

Additionally, a Repeated Measures ANalysis of VAriance (RM-ANOVA)\cite{keselman2001analysis} was conducted, focusing on the within-subject variable of the reproduction method. This analysis revealed a statistically significant main effect for the method variable, as indicated by $F(3,45) = 54.04$, $p<0.001$ and $\eta_p^2=0.782$. Due to this significant effect, a post-hoc test with Bonferroni correction was performed to delve deeper into the findings. Pairwise comparisons of the estimated marginal means showed no significant difference between the reference signal and the spherical array estimate, mean difference of $7.687$ and $p=0.057$. In contrast, significant differences were found between the reference signal and both the synthesized signal and the semi-circular array estimate ($p < 0.001$). Furthermore, comparisons between the synthesized signal and the estimates from the spherical ($p < 0.001$) and semi-circular arrays ($p = 0.011$) also showed significant gaps. Finally, the difference between the spherical and semi-circular array estimates was also significant ($p < 0.001$).

These results suggest that the algorithm effectively enhanced spatial perception by incorporating early room reflections. Performing best with the spherical array and slightly worse with the semi-circular array. These findings demonstrate the effectiveness of the FF-PHALCOR algorithm in improving spatial perception by incorporating estimated early room reflections into the reproduced signals. However, this evaluation was limited to a single room configuration, one source, and one type of stimulus (speech). These constraints may limit the generalizability of the results, and further studies with varying room acoustics, different source types, and multiple stimuli are necessary to validate the findings more broadly.

\section{Conclusion}
This paper presents several key contributions. First, the FF-PHALCOR algorithm was applied to a semi-circular array, including a comprehensive performance analysis, while previous work only studied spherical arrays. Secondly, an understanding was gained of how reflection characteristics, such as delay, amplitude, and spatial density, affect performance. Thirdly, enhancements in the clustering stage minimized both missed detections and false alarms in estimating reflection directions of arrival and delays. Lastly, the algorithm demonstrated its effectiveness in improving spatial perception by incorporating reflection estimates into reproduced room impulse responses.

The results showed that the FF-PHALCOR algorithm performed well with the semi-circular array, although its performance was slightly degraded compared to the spherical array. There was a significant alignment between detection ability and the delay and amplitude of reflections, with false alarms being related to the delay. Addressing clustering issues improved performance, and considering only azimuth angles reduced false alarms in the semi-circular array. In terms of spatial perception, the spherical array showed no significant gap from the reference, while the semi-circular array performed well compared to the anchor, albeit with a significant gap.

One limitation of the method is the reduced performance with fewer microphones, i.e. with the semi-circular array compared to the spherical array. Future work could optimize the algorithm for different array configurations to improve consistency across various arrays. Further research could also investigate the impact of different room shapes and acoustic properties on the algorithm's performance. Additionally, implementing real-time processing and testing in diverse real-world environments would enhance the practical applicability of the FF-PHALCOR algorithm. Finally, a more comprehensive listening test with full RIR is also proposed for future work.

% or


%\appendices

%\section{}

\section*{Acknowledgment}

This work was partially supported by Reality Labs Research @ Meta

\ifCLASSOPTIONcaptionsoff
  \newpage
\fi



\bibliographystyle{IEEEtran}
\bibliography{refs}


\begin{IEEEbiography}[{\includegraphics[width=1in,height=1.25in,clip,keepaspectratio]{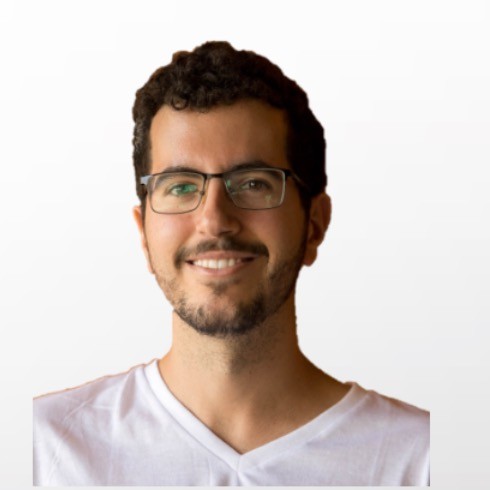}}]{Yogev Hadadi}
Yogev Hadadi received the B.Sc. degree in electrical and computer engineering from Ben-Gurion University of the Negev, Beer-Sheva, Israel, in 2022. He was honored on the Dean's List at Ben-Gurion University of the Negev in 2023. He is currently pursuing his M.Sc. and working in GNSS signal processing.\end{IEEEbiography}

\begin{IEEEbiography}[{\includegraphics[width=1in,height=1.25in,clip,keepaspectratio]{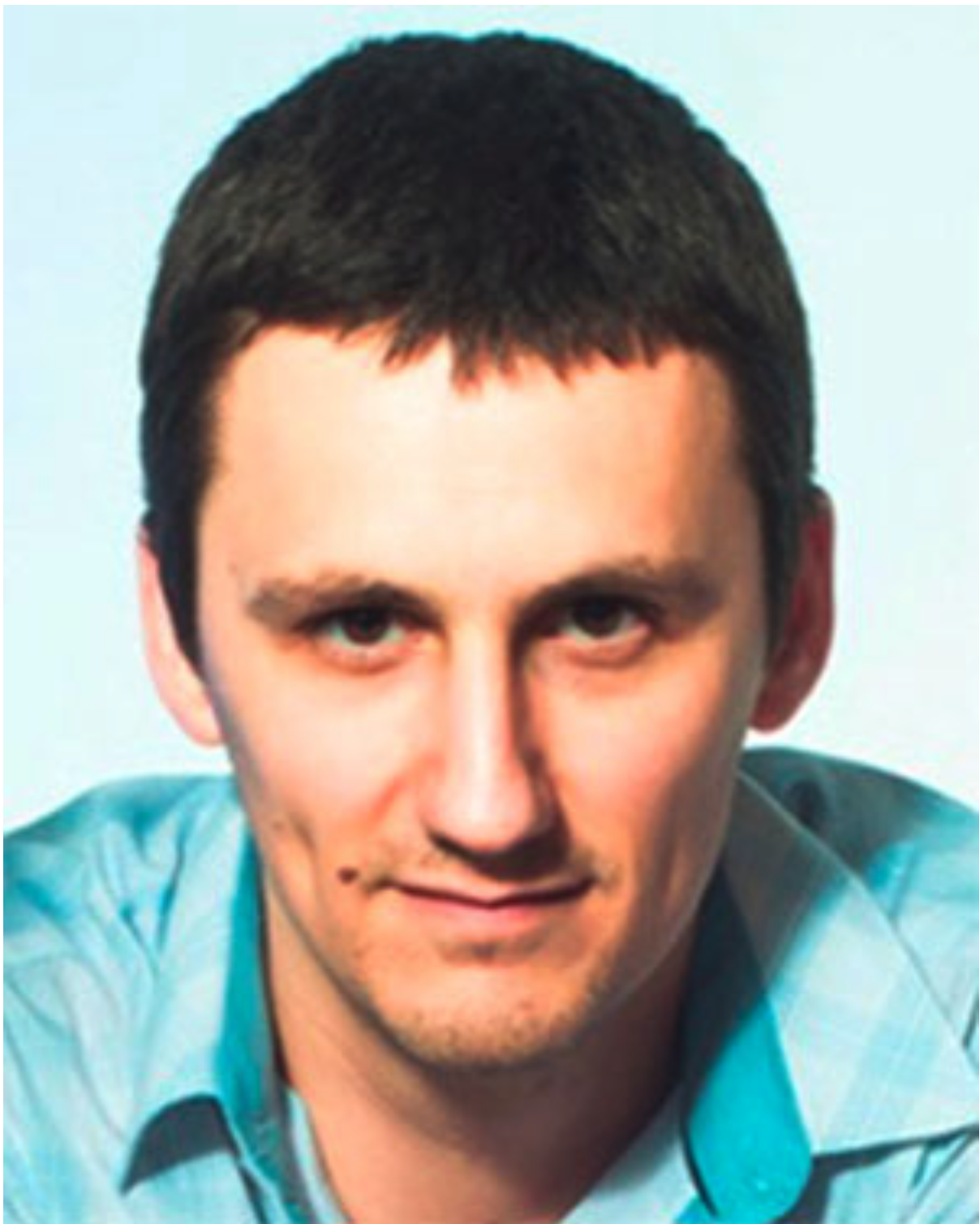}}]{Vladimir Tourbabin}
(Member, IEEE) received the B.Sc. degree (summa cum laude) in materials science and engineering, the M.Sc. (cum laude) and Ph.D. degrees in electrical and computer engineering from the Ben-Gurion University of the Negev, Be’er Sheva, Israel, in 2005, 2011, and 2016, respectively. He is currently a Research Science Manager with Meta, Cambridge, MA, USA. After the graduation, he has joined the General Motors’ Advanced Technical Cen- ter in Israel, to work on microphone array processing solutions for speech recognition. Since 2017, he has
been with Reality Labs Research @ Meta (formerly known as Facebook Reality Labs) working on research and advanced development of audio signal processing technologies for augmented and virtual reality applications.
\end{IEEEbiography}

%\newpage

\begin{IEEEbiography}[{\includegraphics[width=1in,height=1.25in,clip,keepaspectratio]{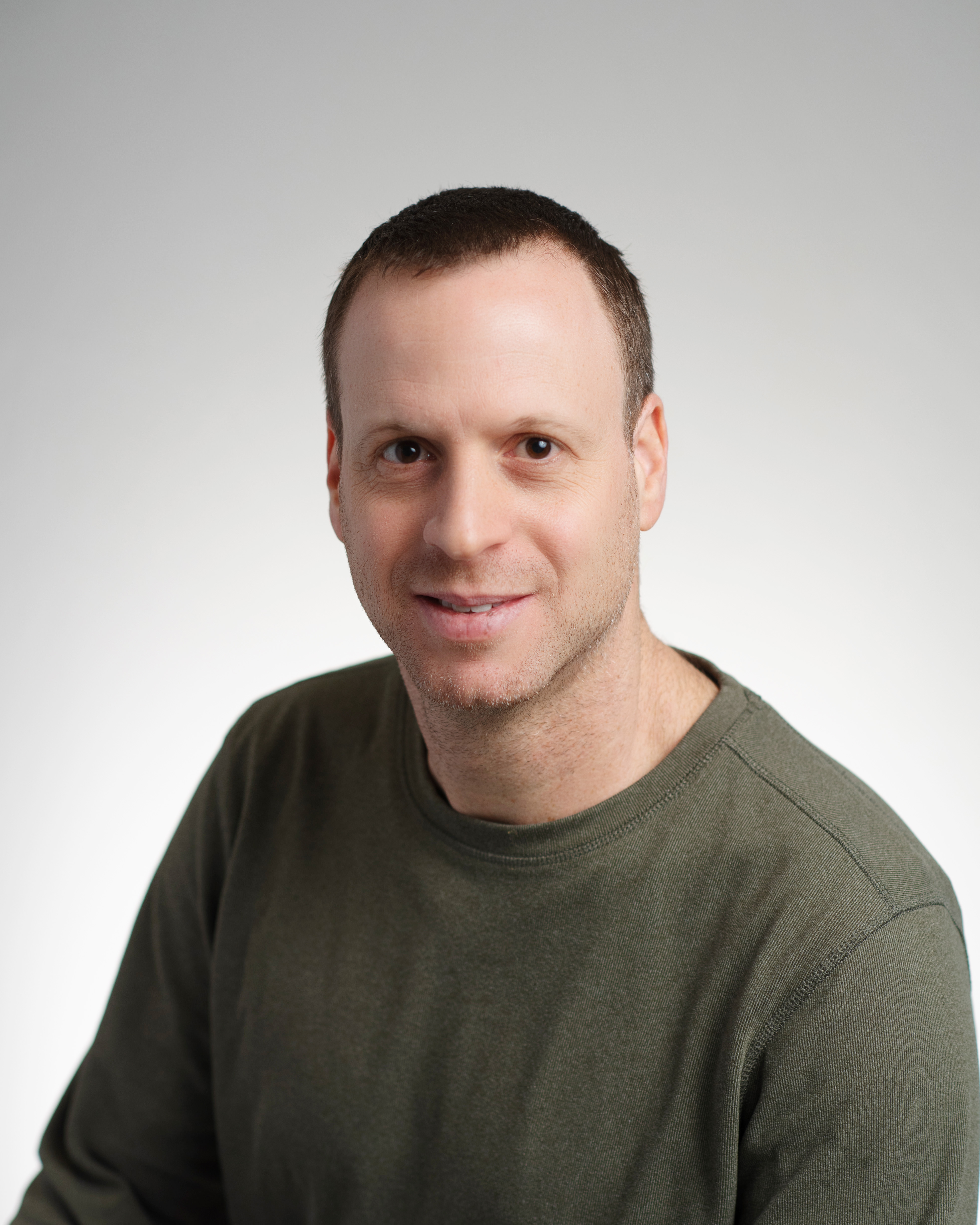}}]{Zamir Ben-Hur}
Zamir Ben-Hur is currently a research scientist at Meta Reality Labs Research, working on spatial audio technologies. He received his B.Sc. (summa cum laude), M.Sc., and Ph.D. degrees in electrical and computer engineering in 2015, 2017, and 2020, respectively, from Ben-Gurion University of the Negev, Beer-Sheva, Israel. His research interests include spatial audio capture, microphone array signal processing and binaural reproduction. 
\end{IEEEbiography}

\begin{IEEEbiography}[{\includegraphics[width=1in,height=1.25in,clip,keepaspectratio]{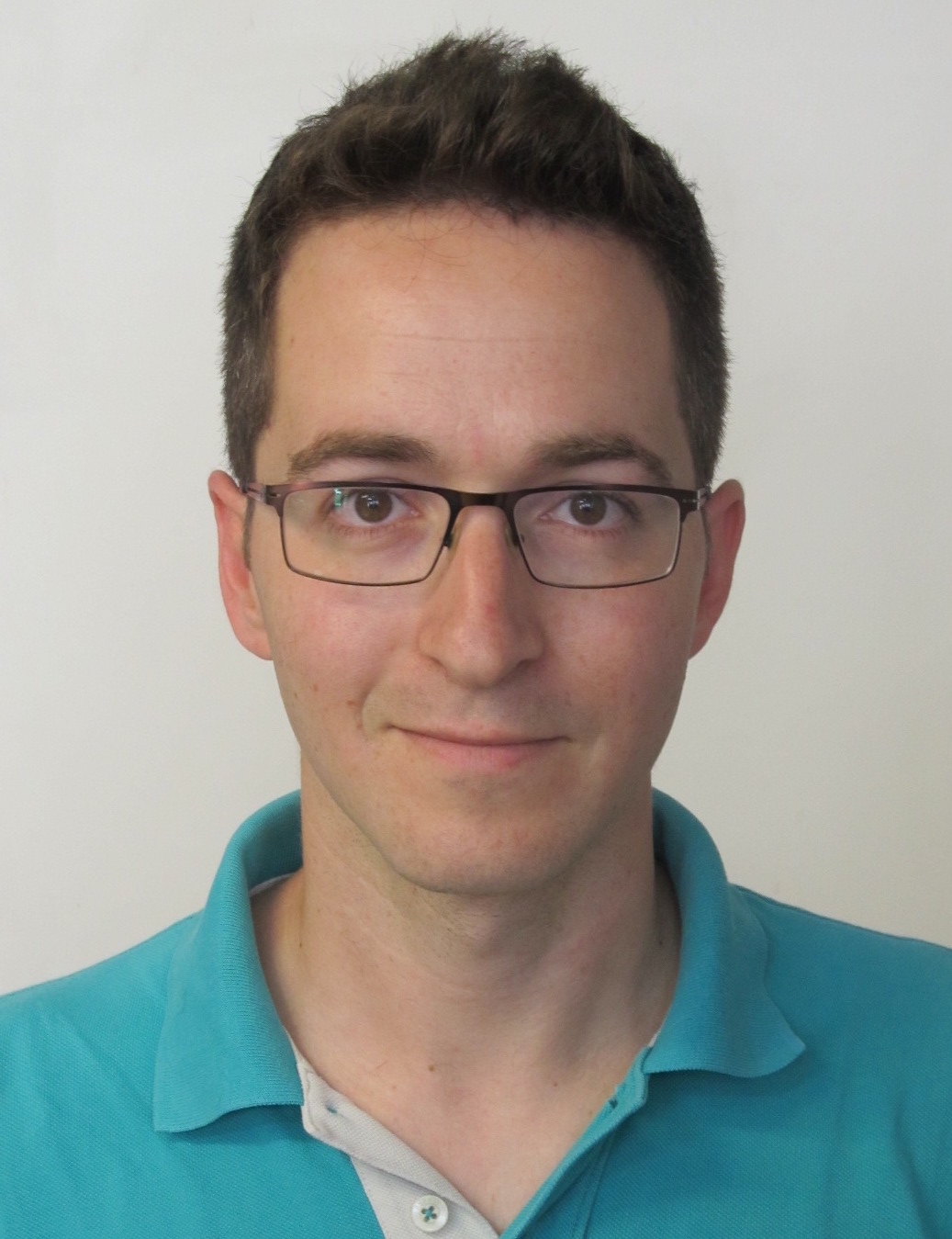}}]{David Lou Alon}
David Lou Alon is a Research Scientist at Meta Reality Labs Research, investigating spatial audio technologies. He received his Ph.D. in electrical engineering from Ben Gurion University (Israel, 2017) in the field of spherical microphone array processing. His research areas include head-related transfer functions, spatial audio capture, binaural reproduction, and headphone equalization for VR and AR applications. 
\end{IEEEbiography}

\begin{IEEEbiography} 
[{\includegraphics[width=1in,height=1.25in,clip,keepaspectratio]{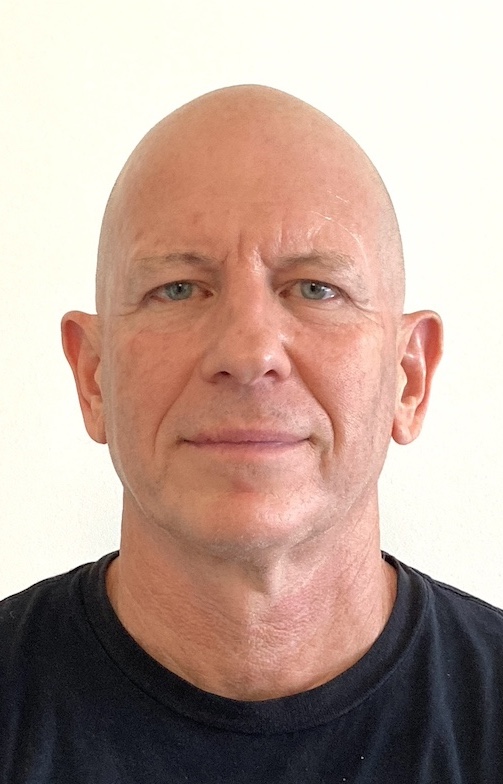}}]{Boaz Rafaely}
Boaz Rafaely (SM’01) received the B.Sc. degree (cum laude) in electrical engineering from Ben-Gurion University, Beer-Sheva, Israel, in 1986; the M.Sc. degree in biomedical engineering from Tel-Aviv University, Israel, in 1994; and the Ph.D. degree from the Institute of Sound and Vibration Research (ISVR), Southampton University, U.K., in 1997. 
At the ISVR, he was appointed Lecturer in 1997 and Senior Lecturer in 2001, working on active control of sound and acoustic signal processing. In 2002, he spent six months as a Visiting Scientist at the Sensory Communication Group, Research Laboratory of Electronics, Massachusetts Institute of Technology (MIT), Cambridge, investigating speech enhancement for hearing aids. He then joined the Department of Electrical and Computer Engineering at Ben-Gurion University as a Senior Lecturer in 2003, and appointed Associate Professor in 2010, and Professor in 2013. He is currently heading the acoustics laboratory, investigating methods for audio signal processing and spatial audio. During 2010-2014 he has served as an associate editor for IEEE Transactions on Audio, Speech and Language Processing, and during 2013-2018 as a member of the IEEE Audio and Acoustic Signal Processing Technical Committee. He also served as an associate editor for IEEE Signal Processing Letters during 2015-2019, IET Signal Processing during 2016-2019, and currently for Acta Acustica. During 2013-2016 he has served as the chair of the Israeli Acoustical Association, and during 2016-2022 as the chair of the Technical Committee on Audio Signal Processing in the European Acoustical Association. He has served as the head of the School of Electrical and Computer Engineering at Ben-Gurion university between 2021-2024. 
Prof. Rafaely was awarded the British Council’s Clore Foundation Scholarship.

\end{IEEEbiography}


%\vfill


\end{document}